\documentclass[12pt]{article}
\usepackage{amsfonts}
\usepackage{amsmath}
\usepackage{amssymb}
\usepackage{theorem}
\usepackage{enumerate}
\usepackage{latexsym}
\usepackage{graphicx}

\newcommand{\bbC}{\mathbb C}
\newcommand{\bbN}{\mathbb N}

\newcommand{\cA}{\mathcal A}
\newcommand{\cB}{\mathcal B}
\newcommand{\cO}{\mathcal O}
\newcommand{\cI}{\mathcal I}
\newcommand{\cJ}{\mathcal J}

\newcommand{\cL}{\mathcal L}

\def\a{\alpha}
\def\b{\beta}
\def\d{\delta}

\def\g{\gamma}

\newtheorem{teo}{Theorem}[section]
\newtheorem{lem}{Lemma}[section]
\newtheorem{tve}{Proposition}[section]

\title{\bf Zhukovsky-Volterra top and quantisation ideals}

\author{A. Mikhailov$^{1}$, T. Skrypnyk$^{1,2}$\\
{\small $^1$  School of Mathematics, University of Leeds, LS2 9JT, UK ,} \\
 {\small  $^2$  Bogolyubov Institute for Theoretical Physics,
Metrologichna st.14-b, 03143,  Kiev, Ukraine }}

\date{ }
\begin{document}
\thispagestyle{empty}
 \maketitle
\begin{abstract}
In this letter, we revisit the quantisation problem for a fundamental model of classical mechanics—the Zhukovsky-Volterra top. We have discovered a four-parametric pencil of compatible Poisson brackets, comprising two quadratic and two linear Poisson brackets. Using the quantisation ideal method, we have identified two distinct quantisations of the Zhukovsky-Volterra top. The first type corresponds to the universal enveloping algebras of $so(3)$, leading to Lie-Poisson brackets in the classical limit. The second type can be regarded as a quantisation of the four-parametric inhomogeneous quadratic Poisson pencil. We discuss the relationships between the quantisations obtained in our paper, Sklyanin's quantisation of the Euler top, and Levin-Olshanetsky-Zotov's quantisation of the Zhukovsky-Volterra top.
\end{abstract}


\newpage
\section{Introduction}
The classical and quantum tops are fundamental models in physics. The anisotropic Zhukovsky-Volterra \cite{Volt, Zhuk} and Euler tops \cite{LL} stand out as the simplest yet non-trivial examples. In the classical case, they describe the motion of a free rigid body in the presence or absence of an external field. In the quantum domain, they characterise the dynamics of an isolated spinning particle, an atom or a nucleus subjected to a constant external field, and contribute to the description of phase transitions in atomic nuclei in the Lipkin-Meshkov-Glick model (see \cite{SkrSIGMA2022} and references therein). Their significance in classical and quantum mechanics motivates  us  to revisit the problem of quantisation using a novel approach.

The classical Zhukovsky-Volterra system \cite{Volt, Zhuk} is a dynamical system in three-dimensional phase space with coordinates $S_{\alpha},\ \alpha \in \overline{1,3}$ (representing rotational momenta or classical spin), described by the following system of three ordinary differential equations:
\begin{equation}\label{CompZhuk0}
\frac{dS_{\alpha}}{dt}= 2(j_{\beta}-j_{\gamma}) S_{\beta} S_{\gamma}+ 2(k_{\beta} S_{\gamma}- k_{\gamma} S_{\beta}),
\end{equation}
where $\alpha,\beta,\gamma$ represent a cyclic permutation of the indices $1,2,3$, $j_{\alpha}$ are parameters of anisotropy (reciprocals of the components of the inertia tensor), and $k_{\alpha}$ are the constant components of the external field.


It is well-known that the Zhukovsky-Volterra (and Euler) top admits a pencil of compatible Lie--Poisson brackets. By extending the phase space with a new coordinate $S_0$, which is a constant of motion of the dynamical system, one can also construct quadratic Poisson brackets on the resulting four-dimensional space.
For the Euler top, the later coincides with the famous Sklyanin algebra \cite{Skl1982}. In the case of the Zhukovsky-Volterra top, it represents a known modification of the Sklyanin algebra \cite{Zotov}.

In this paper, we demonstrate that Zhukovsky-Volterra and Euler tops admit a second inhomogeneous quadratic Poisson bracket. Furthermore, we establish that all four Poisson brackets are mutually compatible in the sense of Magri \cite{Magri}, and define a  family of Poisson brackets:
\begin{equation}\label{pb1}
 \begin{array}{l}
\{S_{\alpha}, S_{\beta}\}_{a,b,c,d}= 2\left((c +d\, j_{\alpha} ) S_{0} S_{\gamma} + (a+ b\,j_{\gamma} )S_{\gamma}+ d \,k_{\gamma}   S_{0}+ b\,k_{\gamma}   \right),\\
\{S_{0}, S_{\alpha}\}_{a,b,c,d}= 2e\, (j_{\beta}-j_{\gamma}) S_{\beta} S_{\gamma} + 2 e\, (k_{\beta} S_{\gamma} - k_{\gamma} S_{\beta}),
 \end{array}
\end{equation}
where $a$, $b$, $c$, $d$ and $e$ are arbitrary parameters of the family (if $e\ne 0$, then by a rescaling of the bracket one can set $e=1$ without a loss of generality).
The existence of the second quadratic structure in these models, along with the compatibility of all four Poisson brackets, seems to be a novel result to the best of our knowledge.  If $e\ne 0$, the variable $S_0$ plays the role of the Hamiltonian for the entire family of the brackets.

We address the quantisation problem through a novel approach based on the concept of quantisation ideal, initially introduced in \cite{Mikhailov2020} and further developed in \cite{CMW22}-\cite{MV}. This approach is tailored for dynamical systems defined on free associative algebras. As a preliminary step, we lift the Zhukovsky-Volterra system to the free algebra $\cA=\bbC\langle \hat{S}_0,\hat{S}_1,\hat{S}_2,\hat{S}_3\rangle$ using the same Lax representation as applied to the matrix-valued system \cite{Olshan}. The resulting system defines a derivation $\frac{d}{dt}\,:\, \cA\rightarrow\cA$ of the algebra. Subsequently, we seek a quantisation ideal $\cJ\subset\cA$, which is an ideal satisfying two conditions:
\begin{enumerate}
 \item[(i)] the ideal $\cJ$ is $\frac{d}{dt}$--stable:\;$\frac{d}{dt}\cJ\subset\cJ$;
 \item[(ii)]  the quotient algebra $\cA/\cJ$ admits a basis of normally ordered monomials.
\end{enumerate}
The quotient algebra  $\cA/\cJ$ is then said to be a quantum algebra for the system. The first condition implies that
$\frac{d}{dt}$ decends to a derivation of the quantum algebra, defining the quantum Zhukovsky-Volterra system
with commutation relations determined by the generators of the quantisation ideal.

As a candidate for a quantisation ideal, we consider the ideal $\cI$ genrated by the set of polynomials:
\[ \begin{array}{ll}
\cI=\langle\hat{S}_{\a}  \hat{S}_{\b} - \hat{S}_{\b}\hat{S}_{\a} +  A_{\g} ( \hat{S}_{0} \hat{S}_{\g} +  \hat{S}_{\g} \hat{S}_{0}) +   K_{\g}\hat{S}_{0}+ M_{\g}\hat{S}_{\g}+N_{\g} , \\
 \hat{S}_{\g}  \hat{S}_{0}  -   \hat{S}_{0}\hat{S}_{\g}+ B_{\g} ( \hat{S}_{\a} \hat{S}_{\b} +  \hat{S}_{\b} \hat{S}_{\a})+    ( {l}_{\a} \hat{S}_{\b} -  {l}_{\b} \hat{S}_{\a}) \,|\,   \a,\b,\g\in \overline{1,3}\rangle.
   \end{array}
\]
These polynomials generalise the commuation relations in the Sklyanin algebra \cite{Skl1982}
and the algebra obtained in \cite{Zotov}. The ideal $\cI$ is parametried by the set of 18 constants $A_\a,B_\a,K_\a,M_\a,N_\a $ and $l_\a,\ \a\in \overline{1,3}$. The fulfillment of conditions (i) and (ii) leads to a system of algebraic equations on these parameters, which we have  solved in the paper to derive the most general quantisation ideal of the form $\cI$.

In the simplest case $A_\g=B_\g=0$ condition (ii) is satisfied due to the Poincare-Birkchoff-Witt Theorem. Ideals satisfying condition (i) almost immideatly give rise to a quantum algebra   isomorphic to  the  universal enveloping algebra given by (\ref{pb1}) with $c=d=e=0$, and  with the central element $\hat{S}_0$.

We treat the generic case $A_\gamma \neq 0, B_\gamma \neq 0$ separately. Firstly, we identify and solve equations for the coefficients of the ideal to fulfill condition (ii) for quadratic and cubic monomials (Theorem \ref{PBWconds}). Condition (i)  imposes   firther constraints on the coefficients of the ideal (Propositions \ref{Heis} and \ref{prop34}). Finally we propose a reparametrisation of the coefficients that resolve all the constraints and is convenient for a classical limit.

Ultimately, we obtain a general quantisation ideal  that leads to commutation relations:
\begin{equation}\label{comrel}
\begin{array}{ll}
   \phantom{.}&
 [\hat{S}_{\a},  \hat{S}_{\b}]= - \frac{  h}{ 1 +  h^2 j_{\g}\bigl(C + (j_{\b}+j_{\a}) D \bigr)}\Bigl((C+  D  j_{\g})   ( \hat{S}_{0} \hat{S}_{\g} +  \hat{S}_{\g} \hat{S}_{0})  +2(A+  B j_{\g} )\hat{S}_{\g}+    2D k_{\g}   \hat{S}_{0}+     2B k_{\g} \Bigr),\\
&
[\hat{S}_{\g}, \hat{S}_{0}] =  -  h\Bigl( {(j_{\a}-j_{\b})} ( \hat{S}_{\a} \hat{S}_{\b} +  \hat{S}_{\b} \hat{S}_{\a})+ 2( {k}_{\a} \hat{S}_{\b} -  {k}_{\b} \hat{S}_{\a})\Bigr) .
  \end{array}
\end{equation}
Here $h=-i \hbar$ can be regarded as a quantisation (Planck) constant.

The five parameters $h,\, A,\,B,\,C,\,D$ defines the quantum algebra. The classical limit corresponds to $h=0$, in which the algebra becomes commutative. We may assume that the parameters $A=A(h),\,B=B(h),\,C=C(h),\,D=D(h)$ are analytic functions of the variable $h$ at $h=0$. They represent a trajectory  to the classical boundary in the space of parameters.  Evaluating the standard classical limit of the commuation relations   (\ref{comrel}) results in the Poisson brackets family (\ref{pb1}), where $a=A(0),\,b=B(0),\,c=C(0),\,d=D(0),\, e=1$.

 The quantum Sklyanin algebra \cite{Skl1982} corresponds to the case of the Euler top ($k_{\a}=0$), with homogeneous quadratic commutation relations having $A=B=0,\, e=1 $ and  certain choices of functions $C(h)$ and $D(h)$ (see  Section  \ref{sklalgex}).  In the quantisation of the Zhukovsky-Volterra by Levin-Olshnetsky-Zotov  \cite{Zotov}, the commutation relations do not have the form   (\ref{comrel}), and they do not satisfy condition (i) for the lifted equation on   $\mathcal{A}$, but satisfy condition (ii). Their quantum system is a deformation of the classical one, i.e. they deform both the commutative algebra and the equation of motion.

The structure of the present paper is as follows: in Section 2, we explore the classical Zhukovsky-Volterra (and Euler) top, its integrals and four compatible Poisson structures. In Section 3, we apply the quantization ideal method and identify a five-dimensional variety of quantizations. Finally, in Section 4, we provide a brief summary and discuss the open problems.

\section{Classical Zhukovsky-Volterra and Euler  tops}
\subsection{Equation of motion and its first integrals}
Motion of a rigid body in a constant external field, known as the classical Zhukovsky-Volterra  top,
can be characterised by the vector of angular momenta $ (S_1,S_2,S_3)\in \mathbb{R}^3$, whose components satisfy  the following system of   ordinary differential equations
 \begin{equation}\label{CompZhuk}
 \frac{d  S_{\a}}{dt}= 2(j_{\b}-j_{\g})  S_{\b} S_{\g}+ 2( k_{\b} S_{\g}- k_{\g} S_{\b}).
 \end{equation}
Here $(k_1,k_2,k_3)$  denotes a vector of  a constant  external field and $j_1,j_2,j_3$ are reciprocals of inertia momentum of the diagonal inertia tensor. In system  (\ref{CompZhuk}) and thereafter we assume that indices
 $(\a,\b,\g)$ represent a cyclic permutation  of the table $(1,2,3)$.

The classical Zhukovsky-Volterra  top (\ref{CompZhuk}) admits two first integrals
\begin{equation}\label{CH}
C=\frac{1}{2}\sum\limits_{\a=1}^3 S^2_{\a},\qquad
 H= \frac{1}{2}\sum\limits_{\a=1}^3 j_{\a}S^2_{\a}+ \sum\limits_{\a=1}^3 k_{\a}S_{\a}.
\end{equation}
When the external field vanishes ($k_{\alpha}=0$), system (\ref{CompZhuk}) reduces to the classical Euler top.

\subsection{Linear Poisson pencil}

It is well known that system (\ref{CompZhuk}) is Hamiltonian
\begin{equation*}
 \frac{d  S_{\a}}{dt}= \{ S_{\a}, H \}_1.
 \end{equation*}
 with respect to the standard   linear Poisson structure on $so^*(3)$:
\begin{equation}\label{so3br}
\{S_{\a}, S_{\b}\}_1=2  S_{\g},
\end{equation}
and the Hamiltonian $H$. It is also Hamiltonian
\begin{equation*}
 \frac{d  S_{\a}}{dt}= \{C, S_{\a}\}'_1.
 \end{equation*}
 with respect to the linear inhomogeneous Poisson structure
\begin{equation}\label{so3br'Zhuk}
\{S_{\a}, S_{\b}\}'_1=2  (j_{\g}S_{\g}+ k_{\g}),
\end{equation}
and the Hamiltonian $C$.

 The functions $C$ and $H$ are Casimir functions of the brackets $\{\ , \ \}_1$ and $\{\ , \ \}'_1$, respectively, i.e.,
 $$\{ C, S_{\a}\}_1=\{ H, S_{\a}\}'_1=0, \quad \a\in\overline{1,3}.$$

The brackets  $\{\ , \ \}_1$ and $\{\ , \ \}'_1$ are compatible in the sense of Magri \cite{Magri},
meaning that a linear combination of the brackets
\begin{equation}\label{linpuch}
\{S_{\a}, S_{\b}\}_{a,b}=2  (a+ j_{\g} b)S_{\g}+   2k_{\g} b,
\end{equation}
with arbitrary constant parameters $a$ and $b$ is a  Poisson bracket. The
Casimir function of this bracket is the following function
$$
C_{a,b}=a C+b H.
$$

\subsection{Extension of the phase space and  quadratic Poisson structures}
Let us now consider  quadratic Poisson structure for the Zhukovsky-Volterra  top.
For this purpose we need to extend  the phase space of the model with a new variable $S_0$, which is a constant of motion ($
\frac{d S_0}{dt}= 0
$)
and a central element of two linear brackets:
 $$\{ S_0, S_{\a}\}_1=0, \quad \{ S_0, S_{\a}\}'_1=0,  \qquad  \a\in\overline{1,3}.$$

In the extended phase space we are looking for  an inhomogeneous quadratic Poisson structure   of the  form:
\begin{subequations}\label{qpbZhuk}
\begin{equation}\label{qpbZhuk1}
\{S_{\a}, S_{\b}\}= 2  a_{\g} S_{0} S_{\g} + 2  K_{\g} S_{0} ,
\end{equation}
\begin{equation}\label{qpbZhuk2}
\{S_{0}, S_{\a}\}= e \bigl( 2(j_{\b}-j_{\g}) S_{\b} S_{\g} + 2 (k_{\b} S_{\g} - k_{\g} S_{\b})\bigr),
\end{equation}
\end{subequations}
where  $a_{\a}$, $K_{\g}$  are some constants.

\begin{tve}
 (i) The  brackets (\ref{qpbZhuk}) satisfy the Jacobi identity iff
\begin{equation}\label{QPBZhukPuch}
 \ a_{\a}=c + j_{\a} d,   \quad K_{\a}=  k_{\a} d , \quad \a\in\overline{1,3}.
 \end{equation}
(ii) The Poisson brackets given  by (\ref{qpbZhuk}) with the structure constants defined by  (\ref{QPBZhukPuch}) are compatible with the pencil of linear-constant Poisson brackets  $ \{\ , \ \}_{a,b}$ defined by (\ref{linpuch}) for any values of  $c$, $d$,  $a$, $b$, $e$.
\end{tve}

 Furthermore, in order for the equation (\ref{qpbZhuk2}) to coincide exactly with the equation of motion we will further impose the normalization condition $e=1$.  This can be always achieved if $e\neq 0$. The  the case $e=0$ can be  effectively reduced to the previously considered linear case.
   Thus we obtained the Poisson brackets:
 \begin{subequations}\label{qpbZhukGen}
\begin{equation}\label{qpbZhukGen1}
\{S_{\a}, S_{\b}\}_{a,b,c,d}= 2\bigl( (c +j_{\a} d) S_{0} S_{\g} +  (a+ j_{\g} b)S_{\g}+     k_{\g} d S_{0}+    k_{\g} b \bigr) ,
\end{equation}
\begin{equation}\label{qpbZhukGen2}
\{S_{0}, S_{\a}\}_{a,b,c,d}=   2(j_{\b}-j_{\g}) S_{\b} S_{\g} + 2 (k_{\b} S_{\g} - k_{\g} S_{\b}),
\end{equation}
\end{subequations}
which depend on four arbitrary parameters $a,b,c,d$.

The above Proposition can be proven by solving equations on the coefficients $a_\g,K_\g$ that are obtained from the Jacobi identity.  We derived the brackets (\ref{qpbZhukGen}) as a classical limit of the general commutation relations (\ref{QComReQuadZhuk5}) outlined in Proposition \ref{prop36}. Consequently, the fulfillment of the Jacobi identity  is guaranteed for any choice of  $a,b,c,d$.

{\it Remark 1.}  Note that   (\ref{qpbZhukGen}) represents a linear combination of two quadratic brackets if we set  $c+d=1$.

{\it Remark 2.} Observe that in the case $d a-b c=0,\ c\ne0,\ d\ne0$, the linear term in the equation  (\ref{qpbZhukGen1}) can be obtained simply by shift of the element $S_{0}$: $S_{0} \rightarrow S_{0}+\nu$, where $\nu=ac^{-1}= bd^{-1} $. 

Applying the method of indeterminate coefficients, we have identified two Casimir functions of the brackets (\ref{qpbZhukGen}):
\begin{subequations}\label{CasZhukGen}
\begin{equation} \label{CasZhukGen1}
C_1=  \sum\limits_{\a=1}^3 (c + j_{\a} d)S_{\a}^2+2 d  \sum\limits_{\a=1}^3 k_{\a}S_{\a}-2(d a-b c) S_0 ,
\end{equation}
\begin{equation} \label{CasZhukGen2}
C_2=  \sum\limits_{\a=1}^3 S_{\a}^2+ d S^2_0+ 2 b S_0.
\end{equation}
\end{subequations}


\section{Quantum  Zhukovsky-Volterra  top and quantisation ideals}

In order to apply the quantisation ideal approach we lift equations of
the classical commutative Zhukovsky--Volterra  top to a free associative
algebra $\mathcal{A}=\bbC\langle \hat{S}_0,\ldots,\hat{S}_3\rangle$ using a Lax
representation, which is similar to the commutative case.

In the simplest case of Lie type ideals the existence of a basis of normally ordered monomials $\cB=\langle S_0^{n_0}S_1^{n_1}S_2^{n_2}S_3^{n_3}\,|\, n_k\in\bbN\rangle $
in the quotient algebra $\cA/\cI$, or
a PBW basis,
is following from the  Poincar\'e--Birkhoff--Witt Theorem. The stability
condition (i) leads to a quantisation that in the classical limit results in
the pencil of compatible linear Poisson brackets (\ref{linpuch}).

For a quadratic
ideal $\cI\subset\cA$, we examine   the conditions emerging from the
requirement of existence of a PBW basis
in the quotient algebra $\cA/\cI$ (condition (ii)) in order  to find equations on the parameters of the ideal.
We focus on the subspaces spanned by monomials
$\cB_N=\langle
\hat{S}_0^{n_0}\hat{S}_1^{n_1}\hat{S}_2^{n_2}\hat{S}_3^{n_3}\,|\,
n_0+n_1+n_2+n_3\leqslant N,\  n_k\in\bbN\rangle $ with $N=2$ and $N=3$.

Subsequently, we   consider the stability condition (i). The quantisation
obtained in this way depends on five parameters, and in the commutative
classical limit results in a four parametric family of compatible Poisson
brackets (\ref{qpbZhukGen}). Finally we compare the results obtained with
Sklyanin's quantisation of the Euler top and Levin--Olshanetsky--Zykov
quantisation of the   Zhukovsky--Volterra  top.

\subsection{Equation of motion on free associative algebra}

In the classical  commutative case the Lax representation for the
Zhukovsky-Volterra model was discovered in  \cite{Fedorov} and studied in
\cite{Zotov}, \cite{SkrypnykJPA2017}. The Lax pair
with matrix valued entries $S_\a$  was  discussed  in   \cite{Olshan}.
Here, we employ the same Lax pair, replacing  matrices $S_\a$ by   elements $\hat{S}_\a$ from the free algebra $\mathcal{A}$:
\begin{equation}\label{Qlaxpair}
 \hat{L}=i\sum\limits_{\a=1}^3 (u_{\a} \hat{S}_{\a} + \frac{k_{\a}}{u_{\a}}) \sigma_{\a}, \quad   \hat{M}=i\sum\limits_{\a=1}^3 u_{\b}u_{\g} \hat{S}_{\a} \sigma_{\a},
\end{equation}
where  $u_{\a}$, $\a\in\overline{1,3}$ are   coordinates of a  point on the elliptic spectral curve
$$
u^2_{1}-u^2_{2}=j_{1}-j_2, \qquad u^2_{2}-u^2_{3}=j_{2}-j_3,$$
and $\sigma_{\a}$   are standard Pauli matrices.
The Lax equation
\begin{equation}\label{QLaxEq}
\frac{d \hat{L} }{dt}=[\hat{L} ,\hat{M} ],
\end{equation}
leads to the dynamical system
\begin{equation}\label{QCompZhuk}
 \frac{d  \hat{S}_{\a}}{dt}= (j_{\b}-j_{\g})  (\hat{S}_{\b} \hat{S}_{\g}+
\hat{S}_{\g} \hat{S}_{\b})+ 2 ( {k}_{\b} \hat{S}_{\g}- {k}_{\g} \hat{S}_{\b}),
\qquad \a\in\overline{1,3},
 \end{equation}
that together with the equations
\begin{equation}\label{QEqS0}
\frac{d \hat{S}_0}{dt}= 0,
\end{equation}
represent a lift of the of the classical commutative Zhukovsky-Volterra
top (\ref{CompZhuk}) to the free algebra $\cA$.

\subsection{Quantisation ideals of Lie type}
Let us at first consider a Lie type ideal $\mathcal{J}$, generated by  polynomials:
\begin{equation} \label{QComRelLinZhuk}
\cJ=\langle f_\g=\hat{S}_{\a}  \hat{S}_{\b} - \hat{S}_{\b}\hat{S}_{\a} -  a_{\g} \hat{S}_{\g} - b_{\g}, \  g_\g=\hat{S}_{0}  \hat{S}_{\g}  -   \hat{S}_{\g}\hat{S}_{0}\,|\,  \a,\b,\g\in \overline{1,3}\rangle,
\end{equation}
where $a_\g,b_\g$ are six arbitrary constants and $\hat{S}_{0}$ is a central element of the algebra $\cA$.
It follws from the Poincare-Birkchoff-Witt theorem that the quotient algebra $\mathcal{A}/\mathcal{J}$ admits a basis of normally ordered monomials $\cB=\langle S_0^{n_0}S_1^{n_1}S_2^{n_2}S_3^{n_3}\,|\, n_k\in\bbN\rangle $, which is refered as a PBW basis.

The ideal (\ref{QComRelLinZhuk}) is a quantisation ideal for system (\ref{QCompZhuk}), (\ref{QEqS0}) if $\cJ$ it is stable with respect to the dynamics.

\begin{tve}
The   ideal $\cJ$ is $\frac{d}{dt}$--stable, i.e. $\frac{d\cJ}{dt}\subset\cJ$
  if and only  if  
\begin{equation}\label{lbcondZhuk}
 a_{ \g}=2 (a + j_{\g} b), \quad b_{\g}= 2k_{\g} b,
\end{equation}
where $a,b$ are arbitrary constants.
\end{tve}

{\em Sketch of the proof:} The stability conditions can be obtained from the requirement that the time derivatives of the ideal generators belong to the ideal. The conditions of stability of the generators $g_\g$ do not impose any constraints on the constants $a_{ \g},b_{ \g}$.  It follows from the conditions $\frac{df_\g}{dt}\subset\cJ,\ \g\in \overline{1,3}$ that
$$
(j_1-j_2)a_3+(j_2-j_3)a_1+(j_3-j_1)a_2=0,\qquad b_\g (j_\a-j_\b)=k_\g (a_\a-a_\b),\quad \a,\b,\g\in \overline{1,3}.
$$
The above implies (\ref{lbcondZhuk}).\hfill $\Box$

The statement of the Proposition    means that  the   quantisation ideal of the Lie type effectively depend   on two parameters
\[
 \cJ_{(a,b)}=\langle \hat{S}_{\a}  \hat{S}_{\b} - \hat{S}_{\b}\hat{S}_{\a} - 2 (a + j_{\g} b) \hat{S}_{\g} -2k_{\g} b, \   \hat{S}_{0}  \hat{S}_{\g}  -   \hat{S}_{\g}\hat{S}_{0}\,|\,  \a,\b,\g\in \overline{1,3}\rangle ,
\]
that in the classical limit reduces to the Poisson pencil (\ref{linpuch}). The center of  the quantum algebra  $\cA/\cJ_{(a,b)}$ is generated by $\hat{S}_0$ and $a\hat{C}+b\hat{H}$, where
\begin{equation}\label{H1}
\hat{C}=\frac{1}{2}\sum\limits_{\a=1}^3 \hat{S}^2_{\a},\qquad \hat{H}=\frac{1}{2}\sum\limits_{\a=1}^3 j_{\a}\hat{S}^2_{\a}+ \sum\limits_{\a=1}^3 k_{\a}\hat{S}_{\a}.
\end{equation}

The specification $a= i \hbar$ and $b=0$ leads to the standard commutation relations
 $$
  [\hat{S}_{\a} , \hat{S}_{\b}]= 2i\hbar  \hat{S}_{\g}, \qquad [\hat{S}_{0} , \hat{S}_{\a}]=0
 $$
for   $so(3)$ quantum systems. On the algebra $\cA/\cJ_{(i\hbar,0)}$ the quantum Zhukovsky-Volterra system   (\ref{QCompZhuk}), (\ref{QEqS0}) can be presented in the Heisenberg from
\begin{equation}\label{EqMot2}
 i \hbar \frac{d  \hat{S}_{\a}}{dt}= [ \hat{S}_{\a}, \hat{H}].
\end{equation}

The center of the quantum algebra  $\cA/\cJ_{(i\hbar,0)}$ is generated by the elements $\hat{S}_{0}$ and $\hat{C}$.
The classical limit in this case results in the Poisson brackets (\ref{so3br}), Hamiltonian $H$ and Casimir element $C$ (\ref{CH}).

The second choice of specification  $a=0$ and $b=i \hbar$  leads to  commutation relations
 $$
  [\hat{S}_{\a} , \hat{S}_{\b}]'=  2i\hbar (  j_{\g} \hat{S}_{\g} + k_{\g}), \qquad [\hat{S}_{0} , \hat{S}_{\a}]'=0
 $$
on the algebra $\cA/\cJ_{(0,i\hbar)}$. Here we use   ``prime'' in $[\cdot,\cdot]'$ to emphasize that the multiplication rules the algebras  $\cA/\cJ_{(i\hbar,0)}$ and $\cA/\cJ_{(0,i\hbar)}$ are different.

In the algebra $\cA/\cJ_{(0,i\hbar)}$ the center is generated by the elements $\hat{S}_{0}, \hat{H}$ (\ref{H1}), and the element $\hat{C}$ becomes the Hamiltonian for the quantum Zhukovsky-Volterra system
\begin{equation}\label{EqMot2'}
 i \hbar \frac{d  \hat{S}_{\a}}{dt}= [\hat{C}, \hat{S}_{\a}]'\, .
\end{equation}
The classical limit in the case of algebra $\cA/\cJ_{(0,i\hbar)}$ results in the Poisson brackets (\ref{so3br'Zhuk}), Hamiltonian $C$ and Casimir element $H$ (\ref{CH}).

\subsection{Quadratic ideals: the PBW condition}

We start with consideration of a quite general ideal $\cI\subset\cA$ genrated
by quadratic polynomials:
\begin{equation} \label{QComReQuadZhuk}
 \begin{array}{ll}
\cI=\langle F_\g=\hat{S}_{\a}  \hat{S}_{\b} - \hat{S}_{\b}\hat{S}_{\a} +
A_{\g} ( \hat{S}_{0} \hat{S}_{\g} +  \hat{S}_{\g} \hat{S}_{0}) +
K_{\g}\hat{S}_{0}+ M_{\g}\hat{S}_{\g}+N_{\g} , \\
 G_\g=\hat{S}_{\g}  \hat{S}_{0}  -   \hat{S}_{0}\hat{S}_{\g}+ B_{\g} (
\hat{S}_{\a} \hat{S}_{\b} +  \hat{S}_{\b} \hat{S}_{\a})+    ( {l}_{\a}
\hat{S}_{\b} -  {l}_{\b} \hat{S}_{\a}) \,|\,   \a,\b,\g\in
\overline{1,3}\rangle.
   \end{array}
\end{equation}
These polynomials generalise the commutation relations in the Sklyanin algebra
\cite{Skl1982}
and the algebra obtained in \cite{Zotov}.

The first problem is to find conditions on 18 parameters
$A_\a,B_\a,K_\a,M_\a,N_\a $ and $l_\a,\ \a\in \overline{1,3}$ which guarantee
the existence of the normally ordered monomial basis in the  subspaces
of quadratic and cubic polynomials in   the quotient algebra $\cA/\cI$.

\begin{teo}\label{PBWconds}
\begin{enumerate}
 \item
 Quadratic polynomials in variables $\hat{S}_\a,\ \a\in\{0,1,2,3\}$ admit normal ordering, modulo the ideal $\cI$, i.e. a unique representation in the monomial basis
$\cB_2$, iff
\begin{equation}\label{A2cond}
  A_{1} B_{1} \neq -1,\quad A_{2} B_{2} \neq  1,\quad A_{3} B_{3} \neq
- 1.
\end{equation}
\item Cubic polynomials   admit normal ordering, modulo the ideal $\cI$, if conditions (\ref{A2cond}) satisfied and
\begin{eqnarray}\label{QCondZhuk1}
&&\sum\limits_{\g=1}^3 A_{\g} B_{\g} + \prod\limits_{\g=1}^3 A_{\g} B_{\g}=0,
\\ \label{QCondZhuk2}
&&B_1+B_2+B_3 =0,\\
\label{QCondZhuk3}
K_{\a} &=&  \frac{(A_{\b}-A_{\g} +  B_{\a} A_{\b} A_{\g} )}{B_{\a}(1+ A_{\b}
B_{\b} A_{\g} B_{\g}) }   l_{\a},\\ \label{QCondZhuk4}
M_{\a} &=& 2\nu A_{\a}- \mu \bigl(\frac{3+A_{\b}B_{\b}-A_{\g}B_{\g}+ A_{\b} B_{\b}
A_{\g} B_{\g}}{ 1+ A_{\b} B_{\b} A_{\g} B_{\g}}\bigr), \\ \label{QCondZhuk5}
N_{\a} &=& {\nu} K_{\a}, \qquad  \a\in\overline{1,3},
\end{eqnarray}
\begin{eqnarray}
 &&B_\a\ne 0 ,\quad A_\a^2 B_\a^2 \neq 1,\qquad \a\in\overline{1,3},  \label{A3last0}  \\
&&\left(1+A_1 B_1-A_2 B_2-A_1 A_2 B_2 B_1\right)^2+16 A_1 A_2 B_1 B_2\neq 0,\label{A3last}
\end{eqnarray}
where $\mu$, $\nu$ are arbitrary  parameters, and indices $\a$, $\b$, $\g$ are
cyclic permutation of the set $1,2,3$.
\end{enumerate}
\end{teo}

{\it Sketch of the Proof.}
  (1.) We regard  $F_\g=0,\  G_\g=0, \g\in\{1,2,3\}$ as a
system of six linear  equations with respect to the quadratic monomials
$\hat{S}_{i}\hat{S}_{j},\ i>j$ which are not normally ordered.   This system admits a unique solution if
and only if the conditions (\ref{A2cond}) are satisfied. Its solution enables
us to span any polynomial of degree less or equal to two in the basis  $\cB_2$ of the
normally ordered monomials, modulo the ideal $\cJ$.

(2) The set of possible $64$ cubic monomials contains $20$ normally ordered monomials. The rest $44$ unordered monomials
can be  expressed in the basis $\cB_3$ solving the system    of $48$ polynomial equations
$$\hat{S}_\b F_\a=0,\quad  F_\a \hat{S}_\b=0,\quad \hat{S}_\b G_\a=0,\quad G_\a \hat{S}_\b=0,\quad \a \in\{ 1,2,3\},\ \b\in\{0,1,2,3\}.
$$
The solution of the above system enables one to represent any cubic polynomial in $\cA$ in the basis $\cB_3$ uniquely, modulo the ideal $\cI$.
The resolvability conditions for this overdetermined system of linear equations leads to (\ref{QCondZhuk1})-(\ref{A3last}).
 \hfill $\Box$


{\it Remark 3.} The quantum ideals and PBW conditions for the  quadratic structures of the quantum Euler top are obtained  by  putting $l_{\a}=0$, $K_{\a}=0$, $N_{\a}=0$, $\a\in\overline{1,3}$ in  the  formulae above.


\subsection{Stability of the quadratic ideal}
\subsubsection{Dynamical stability of the ideal and projective parametrisation}

We will denote $\cJ$ the ideal $\cI$ (\ref{QComReQuadZhuk}), whose parameters satisfy conditions (\ref{A2cond})-(\ref{A3last}) (Theorem \ref{PBWconds}). The stability of the ideal with respect to the dynamics (\ref{QCompZhuk}), (\ref{QEqS0}) impose further constraints.

\begin{tve}\label{Heis}
The ideal $\cJ$   is $\frac{d}{dt}$--stable iff
\begin{equation}\label{B2}
B_{\a}=h (j_{\b}-j_{\g}), \quad l_{\a}= 2 h k_{\a},   
\end{equation}
where   $h$ is an arbitrary constant.
\end{tve}

There is a convenient parametrisation of parameters $A_\a$, satisfying   (\ref{QCondZhuk1}) with $B_\a$ satisfying $(\ref{B2})$.

\begin{tve}\label{prop34}
The coefficients $A_{\a}$ satisfying condition (\ref{QCondZhuk1}) with the  constants $B_{\a}$   defined by  (\ref{B2})
can be parametrized as follows:
\begin{equation}\label{A2}
A_{\a}=\frac{1}{ h  J_{\a}}\frac{J_{\b}-J_{\g}}{j_{\b}-j_{\g}}, \qquad \a\in\overline{1,3},
\end{equation}
where $J_{\a}$  satisfy the following   inequalities:
\begin{equation}\label{ineq1}
 J_{\a}\neq 0, \quad J_{\a}\neq J_{\b}+J_{\g}, \qquad \a\in\overline{1,3},
 \end{equation}
\begin{equation}\label{ineq2}
\left(J_1+J_2-J_3\right){}^4+16 J_1 J_2 \left(J_1-J_3\right)
\left(J_2-J_3\right)\neq 0
\end{equation}
 and  are arbitrary otherwise.
\end{tve}

The statement of Proposition \ref{prop34} can be checked by a direct substitution of (\ref{B2}), (\ref{A2}) in (\ref{QCondZhuk1}).
Conditions (\ref{ineq1}), (\ref{ineq2}) represent inequalities (\ref{A3last0}),(\ref{A3last}), where $A_\a$ is given by (\ref{A2}).

{\it Remark 4.} Observe that there exists another then (\ref{A2}) parametrization  of $A_{\a}$, namely:
\begin{equation}\label{A2'}
A_{\a}=-\frac{1}{ h  \tilde{J}_{\a}}\frac{\tilde{J}_{\b}-\tilde{J}_{\g}}{j_{\b}-j_{\g}}, \qquad \a\in\overline{1,3},
\end{equation}
The parametrizations (\ref{A2}), (\ref{A2'}) are equivalent. The equivalence is achieved by an invertible map \cite{ZotPriv}:
$$\tilde{J}_{\a}= J_{\a} (J_{\a}-J_{\b}-J_{\g}).$$
The structure constants $K_{\d}$, as it follows from the formula  (\ref{QCondZhuk3}) and the above  form of $A_{\a}$, $B_{\a}$, $l_{\a}$,  are:
\begin{equation}\label{K2}
K_{\d}=-\frac{2}{h}\frac{  k_{\d}}{ J_{\d}}\frac{\sum\limits_{\a=1}^3 j_{\a}(J_{\b}-J_{\g})}{ \prod\limits_{\a=1}^3(j_{\b}-j_{\g})}, \qquad \d\in\overline{1,3}.
\end{equation}

The structure constants $M_{\a}$, as it follows from the formula  (\ref{QCondZhuk4}) are the following:
\begin{equation}\label{M2}
 M_{\a}= \frac{ 2 \nu}{ h  J_{\a}}\frac{J_{\b}-J_{\g}}{j_{\b}-j_{\g}} +  \mu \frac{(J_1+J_2+J_3)}{J_{\a}}, \qquad \a\in\overline{1,3}.
\end{equation}
In the result the generators of the ideal $\mathcal{J}$ acquire the following explicit form:
\begin{subequations} \label{QComReQuadZhuk'}
\begin{multline}\label{QComRelQuadZhuk1'}
  \hat{S}_{\a}  \hat{S}_{\b} - \hat{S}_{\b}\hat{S}_{\a} +  \frac{1}{ h  J_{\g}}\frac{J_{\a}-J_{\b}}{j_{\a}-j_{\b}} \bigl( (\hat{S}_{0} + \nu) \hat{S}_{\g} +  \hat{S}_{\g} (\hat{S}_{0}+ \nu)\bigr) +   K_{\g} (\hat{S}_{0}+ \nu)+ \frac{   \mu \sum\limits_{\a=1}^3 J_{\a}}{J_{\g}} \hat{S}_{\g},
\end{multline}
\begin{equation}\label{QComReQuadZhuk2'}
  \hat{S}_{\g}  \hat{S}_{0}  -   \hat{S}_{0}\hat{S}_{\g}+ h\bigl((j_{\a}-j_{\b}) ( \hat{S}_{\a} \hat{S}_{\b} +  \hat{S}_{\b} \hat{S}_{\a})+    2( {k}_{\a} \hat{S}_{\b} -  {k}_{\b} \hat{S}_{\a})\bigr), \qquad \g\in \overline{1,3},
\end{equation}
\end{subequations}
where the indices  $\a,\b,\g$ constitute the cyclic permutations of the indices $1,2,3$  and the ideal parameters
$(J_1:J_2:J_3,n,m,h)$ belong to the space $\mathbb{C}P^2\times \mathbb{C}^3$.

\subsubsection{The Casimir elements}
Let us now describe the Casimir  elements of the algebra (\ref{QComReQuadZhuk})  with the structure constants (\ref{B2})-(\ref{K2}).

\begin{tve}\label{qCas}
The following elements:
\begin{multline}\label{Cas1}
\hat{C_1}= - \frac{  \prod\limits_{\a=1}^3(J_{\b}-J_{\g})}{ \prod\limits_{\a=1}^3(j_{\b}-j_{\g}) }\frac{ (\hat{S}_{0}+\nu)^2}{ h^4 \prod\limits_{\d=1}^3 J_{\d}}  +   \sum\limits_{\a=1}^3  \frac{J_{\b}-J_{\g}}{j_{\b}-j_{\g}}  \frac{\hat{S}^2_{\a}}{ h^2 J_{\a}}  + \sum\limits_{\a=1}^3 \frac{K_{\a} }{h}\hat{S}_{\a} 
-   \mu\frac{  (\sum\limits_{\a=1}^3 J_{\a})(\sum\limits_{\a=1}^3 j_{\a}J_{\a}(J_{\b}-J_{\g})) }{ h^3 \prod\limits_{\d=1}^3 J_{\d} \prod\limits_{\a=1}^3(j_{\b}-j_{\g})} \hat{S}_{0},
\end{multline}
and
\begin{multline}\label{Cas2}
\hat{C_2}=  -\sum\limits_{\a=1}^3 \frac{1}{h^2}\frac{J_{\b}-J_{\g}}{j_{\b}-j_{\g}} (J_{\a}-J_{\b}-J_{\g})\hat{S}^2_{\a} +\\ 2  \sum\limits_{\a=1}^3  \frac{k_{\a}}{h^2} \frac{j_{\a} (J_{\b}-J_{\g})(J_{\a}-J_{\b}-J_{\g}) -j_{\b} (J_{\g}-J_{\a}) J_{\g} - j_{\g}(J_{\a}-J_{\b}) J_{\b}}{(j_{1}-j_{2})(j_{3}-j_{1})(j_2-j_3)}\hat{S}_{\a} -  \frac{\mu (\sum\limits_{\a=1}^3 J_{\a})(\sum\limits_{\a=1}^3 j_{\a}(J_{\b}-J_{\g})) }{h^3  \prod\limits_{\a=1}^3( j_{\b}-j_{\g} )} \hat{S}_{0}
\end{multline}
are central elements of the algebra  of the algebra (\ref{QComReQuadZhuk})  with the structure constants (\ref{B2})-(\ref{K2}).
\end{tve}

{\it Idea of the Proof.} The Casimir elements are  found using the method of the indeterminate coefficients in the assumption of the linear-quadratic form of the Casimirs.

{\it Remark 5.}  
From  (\ref{QComReQuadZhuk2'}) it follows that   Heisenberg equation of motion with respect to  $\hat{S}_{0}$:
\begin{equation*}
 i\hbar \frac{d  \hat{S}_{\g}}{dt}= [  \hat{S}_{\g},  \hat{S}_{0}],  \quad \g\in \overline{1,3}
\end{equation*}
 coincides --- on the quotient algebra ---
 with the dynamical equations  (\ref{QCompZhuk})
if and only if  $h=-i\hbar$.

\subsection{The variety of quantum algebras and the classical limit}
\subsubsection{Affine re-parametrisation of the ideal}
In this subsection we   give another parametrisation of the ideal that yields a  quantum analogue of the Poisson pencil structure. It is based on  the observation that a simultaneous re-scaling of the parameters  $ J_\a\to\hat J_\a= QJ_\a,\ Q\ne 0$  does not affect the ideal generated by the polynomials (\ref{QComReQuadZhuk'}).

\begin{lem}\label{LemCD} Let  $\sum\limits_{\a=1}^3 J_{\a} j_{\b} j_{\g} (j_{\b}-j_{\g})\neq 0$. Then up to projective equivalence $\hat{J}_\a=QJ_\a$ the    structure constants $\hat J_{\a}$ can be parametrised as follows:
\begin{equation}\label{Jpencil}
\hat{J}_{\a} 
=1 + h^2 j_{\a}\Bigl(C + (j_{\b}+j_{\g}) D \Bigr), \qquad \a\in\overline{1,3}.
\end{equation}
where  $C$ and $D$ are arbitrary complex parameters.
\end{lem}

{\it Proof.}
The sysetem of three equations on variables $C,D$ and $Q$
\[
 Q J_{\a} 
=1 + h^2 j_{\a}\Bigl(C + (j_{\b}+j_{\g}) D \Bigr), \qquad \a\in\overline{1,3}
\]
admits a unique solution
\begin{equation}\label{C-D}
C= \sum\limits_{\a=1}^3\frac{{J}_{\a} j_{\a}(j_{\b}-j_{\g})}{h^2\sum\limits_{\a=1}^3 J_{\a} j_{\b} j_{\g} (j_{\b}-j_{\g})}, \quad   D= - \frac{ \sum\limits_{\a=1}^3 {J}_{\a} (j_{\b}-j_{\g})}{h^2\sum\limits_{\a=1}^3 J_{\a} j_{\b} j_{\g} (j_{\b}-j_{\g})},
\end{equation}
and
$$
Q=\frac{\prod\limits_{\a=1}^3(j_{\b}-j_{\g})}{\sum\limits_{\a=1}^3 J_{\a} j_{\b} j_{\g} (j_{\b}-j_{\g})}.
$$
\hfill $\Box$

{\em Remark 6.} In terms of the paramters $C,D$ the inequalities (\ref{ineq1}), (\ref{ineq2}) take the form
\begin{equation}\label{inequal}
 \begin{array}{l}
 \left(h^2 \left(C \left(j_1+j_2-j_3\right)+2 D j_1
j_2\right)+1\right){}^4+\\16 h^4 \left(j_1-j_3\right) \left(j_2-j_3\right)
\left(C+D j_1\right) \left(C+D j_2\right) \left(h^2 j_2 \left(C+D
\left(j_1+j_3\right)\right)+1\right) \left(h^2 j_1 \left(C+D
\left(j_2+j_3\right)\right)+1\right)\neq 0,\\ \\
1+h^2 \left(C \left(j_\a+j_\b-j_\g\right)+2 D j_\a j_\b\right)\neq 0,\quad
1+h^2 j_\a \left(C+D \left( j_\b+j_\g\right)\right) \neq 0,\quad
\a\in\overline{1,3}.
 \end{array}
\end{equation}

The following Proposition  is the main result of the present article:
\begin{tve}\label{prop36}
A   quantisation ideal of the form  (\ref{QComReQuadZhuk}) for the Zhukovsky--Volterra top (\ref{QCompZhuk}), (\ref{QEqS0})  leads to a quadratic  quantum algebra with the  commutation relations:
\begin{subequations} \label{QComReQuadZhuk5}
\begin{equation}
 [\hat{S}_{\a},  \hat{S}_{\b}]= - \frac{ 2h}{ 1 + h^2 j_{\g}\bigl(C + (j_{\b}+j_{\a}) D \bigr)}\Bigl((C+  D  j_{\g})  \frac{( \hat{S}_{0} \hat{S}_{\g} +  \hat{S}_{\g} \hat{S}_{0})}{2} +(A+  B j_{\g} )\hat{S}_{\g}+    D k_{\g}   \hat{S}_{0}+     B k_{\g} \Bigr),
\end{equation}
\begin{equation}
[\hat{S}_{\g}, \hat{S}_{0}] =  - h\Bigl( {(j_{\a}-j_{\b})} ( \hat{S}_{\a} \hat{S}_{\b} +  \hat{S}_{\b} \hat{S}_{\a})+ 2( {k}_{\a} \hat{S}_{\b} -  {k}_{\b} \hat{S}_{\a})\Bigr) , \qquad \g\in \overline{1,3},
\end{equation}
\end{subequations}
where   parameters $h,\,  A,\, B,\, C,\, D$ satisfy the inequalities (\ref{inequal}) and arbitrary  otherwise.
\end{tve}

{\it Remark 7.} The parameters $\nu$, $\mu$  are related with the constants  $A,\, B,\,  C,\, D,\,  h$  as follows:
\begin{equation}\label{A-B} \nu D=B
   , \quad 2 \nu h C+\mu \Bigl( 3 +h^2\bigl( 2 D \sum\limits_{\a=1}^3  j_{\b}j_{\g}  +  C \sum\limits_{\a=1}^3  j_{\a} \bigr)\Bigr)=2hA  .
\end{equation}

\subsubsection{The classical limit}
Let us now assume  that
the functions $A$, $B$, $C$, $D$ are analytical functions of $h$:
\begin{equation}\label{ABCD}
A=a+  \cO(h ), \quad B=b+  \cO(h ), \quad C=c +  \cO(h ), \quad D=d+  \cO(h ).
\end{equation}
Under such the assumption the quantum algebra  (\ref{QComReQuadZhuk5}) is a  quantum deformation of the classical  inhomogeneous quadratic Poisson algebra with the Poisson brackets (\ref{qpbZhukGen}) labeled by four parameters $a$, $b$, $c$, $d$. Indeed,  using  (\ref{QComReQuadZhuk5}) and the expansions (\ref{ABCD}),  it is easy to see that in the limit $h\rightarrow 0$  the right-hand-side of (\ref{QComReQuadZhuk5}) is exactly $\{\ ,\ \}_{a,b,c,d}$ multiplied by  $-h$ ( i.e. by $i\hbar$).  The inequalities (\ref{inequal}) are obviously satisfied in the neighbourhood of $h=0$.

In terms of parameters $A$, $B$, $C$, $D$, the central elements (\ref{Cas1'})-(\ref{Cas2'}) of the quantum algebra take the form:
\begin{multline}\label{Cas1'}
\hat{C_1}=  -\frac{h^2 \prod\limits_{\a=1}^3(C+D j_{\a})   (D\hat{S}^2_{0}+ 2 B \hat{S}_0) }{ D \prod\limits_{\a=1}^3 (1+h^2j_{\a}(C+(j_{\b}+j_{\g}) D))} + \sum\limits_{\a=1}^3 \frac{  (C+D j_{\a})   \hat{S}_{\a}^2+ 2  D k_{\a}   \hat{S}_{\a} }{(1+ h^2 j_{\a}(C+(j_{\b}+j_{\g}) D))} -  
\\ - 2  (A D- C B)   \frac{  (D+h^2(C^2+C D \sum\limits_{\a=1}^3  j_{\a} +D^2 \sum\limits_{\a=1}^3  j_{\b}j_{\g})  ) }{ D \prod\limits_{\a=1}^3 (1+h^2j_{\a}(C+(j_{\b}+j_{\g}) D))}  \hat{S}_{0},   
\end{multline}
\begin{multline}\label{Cas2'}
\hat{C_2}=    \sum\limits_{\a=1}^3   (C+D j_{\a})(1+ h^2 ( C (j_{\b}+j_{\g}-j_{\a}) +  D j_{\b}j_{\g}))  \hat{S}_{\a}^2 + 2  \sum\limits_{\a=1}^3   (D+h^2 (C^2+ D C j_{\a}- D^2 j_{\b}j_{\g}) ) k_{\a}  \hat{S}_{\a}\\ -   2  (A D- C B)\hat{S}_{0} \, .  
\end{multline}


The  classical limit of the central elements $\hat{C_1}$ and $\hat{C_2}$ yield Casimir elements (\ref{CasZhukGen}) of the quadratic Poisson bracket  (\ref{qpbZhukGen}):
\begin{equation}
C_1=\lim\limits_{\hbar \rightarrow 0} {\hat{C}_1}=  \lim\limits_{\hbar \rightarrow 0} {\hat{C}_2} =  \sum\limits_{\a=1}^3 (c+dj_{\a})S_{\a}^2+2 d  \sum\limits_{\a=1}^3 k_{\a}S_{\a}-2(a d-b c) S_0
\end{equation}
\begin{equation}
C_2=\lim\limits_{\hbar \rightarrow 0}\frac{1}{h^2} \frac{D}{\prod\limits_{\a=1}^3 (C+ D j_{\a})}\Bigl( {\hat{C}_2 }-   \frac{  D \prod\limits_{\a=1}^3 (1+h^2j_{\a}(C+(j_{\b}+j_{\g}) D))}{ (D+h^2(C^2+C D \sum\limits_{\a=1}^3  j_{\a} +D^2 \sum\limits_{\a=1}^3  j_{\b}j_{\g})  ) } {\hat{C}_1} \Bigr)=  d S^2_0+ \sum\limits_{\a=1}^3 S_{\a}^2+2b S_0.
\end{equation}

\subsection{Comparison with the existing algebras.}
\subsubsection{Sklyanin algebra}\label{sklalgex}
The quantum Sklyanin algebra \cite{Skl1982} corresponds to the case of the purely quadratic structure of quantum anisotropic  Euler's top, i.e to the case $\mu=\nu=0$ and $k_{\a}=0$, $\a\in\overline{1,3}$. The corresponding ideal generators  (\ref{QComReQuadZhuk'}) are simplified to  the following form:
\begin{subequations} \label{SklAl}
\begin{equation}\label{SklAlg1}
F_{\g}= \hat{S}_{\a}  \hat{S}_{\b} - \hat{S}_{\b}\hat{S}_{\a} +  \frac{1}{  h J_{\g}}\frac{J_{\a}-J_{\b}}{j_{\a}-j_{\b}} \bigl( \hat{S}_{0}  \hat{S}_{\g} +  \hat{S}_{\g} \hat{S}_{0}\bigr), \qquad \g\in \overline{1,3},
\end{equation}
\begin{equation}\label{SklAlg2}
G_{\g}= \hat{S}_{\g}  \hat{S}_{0}  -   \hat{S}_{0}\hat{S}_{\g}+ h(j_{\a}-j_{\b}) ( \hat{S}_{\a} \hat{S}_{\b} +  \hat{S}_{\b} \hat{S}_{\a}),
\end{equation}
\end{subequations}
The algebra with commutation relations (\ref{SklAl}) is equivalent to the Sklyanin algebra obtained from the quantum  group considerations, where
\begin{equation}\label{jskl}
J_{\a}=  \frac{  1 + 2 j_{\a}h^2+((j_{\b}+j_{\g}) j_{\a}- j_{\b} j_{\g} )h^4}{(1+h^2 j_{\a})}, \qquad \a\in \overline{1,3}.
\end{equation}
The above expression for $J_\a$ coincides with the one obtained by Sklyanin
after the re-parametrisation  $h^2=1/\wp(i\hbar)$, where
$\wp$ is a Weierstrass elliptic function with $g_2=- 4(j_1j_2+j_2j_3+j_3
j_1),\ g_3=- 4 j_1j_2j_3$, assuming $j_1+j_2+j_3=0$.

It follows from Lemma \ref{LemCD} that corresponding functions $C=C(h)$ and $D=D(h)$ are :
\begin{equation*}
C = \frac{1-   h^4 \sum\limits_{\a=1}^3 j_{\b} j_{\g}-2  h^6 j_1 j_2 j_3}{1  +
h^6 ( \sum\limits_{\a=1}^3 j^3_{\a} + j_1 j_2 j_3) -   h^8\sum\limits_{\a=1}^3 j^2_{\b} j^2_{\g} },
\end{equation*}
\begin{equation*}
D = \frac{h^2(3 +  h^4 \sum\limits_{\a=1}^3 j_{\b} j_{\g})}{1  +
h^6 ( \sum\limits_{\a=1}^3 j^3_{\a}+ j_1 j_2 j_3) -   h^8\sum\limits_{\a=1}^3 j^2_{\b} j^2_{\g} }.
\end{equation*}
In this case $c=1$, $d=0$, i.e. in the classical limit we obtain the  first quadratic (Sklyanin) brackets.

\subsubsection{Algebra of Levin-Olshanetsky-Zotov} \label{zotalgex}
Levin, Olshanetsky and Zotov proposed a quantisation of the Zhukovsky-Volterra top ($k_{\a}\neq 0$)  \cite{Zotov}. Basing on the reflection equation algebra they found a quantum algebra  defined by the following ideal:
\begin{equation}\label{LOZideal0}
\cJ_{LOZ}=\langle \hat{S}_{\a}  \hat{S}_{\b} - \hat{S}_{\b}\hat{S}_{\a} -
 i ( \hat{S}_{0} \hat{S}_{\g} +  \hat{S}_{\g} \hat{S}_{0}), \quad
  \hat{S}_{\g}  \hat{S}_{0}  -   \hat{S}_{0}\hat{S}_{\g}+ i \frac{ (J_{\a}-J_{\b})}{J_{\g}} (
\hat{S}_{\a} \hat{S}_{\b} +  \hat{S}_{\b} \hat{S}_{\a})+   \frac{ i }{J_{\g}} ( {l}_{\a}
\hat{S}_{\b} -  {l}_{\b} \hat{S}_{\a}) \rangle .
\end{equation}
The re-scaling of the variables $$\hat{S}_0\rightarrow  i h {\hat{S}_0},\quad\hat{S}_{\a}\rightarrow \sqrt{J_{\a} J_{\b}} \hat{S}_{\a},\quad{l}_{\a}\rightarrow h \sqrt{J_{\a} J_{\b}} {l}_{\a},\quad\a\in \overline{1,3}$$
transforms the ideal (\ref{LOZideal0}) to the form
\begin{equation}\label{LOZideal}
\cJ=\langle \hat{S}_{\a}  \hat{S}_{\b} - \hat{S}_{\b}\hat{S}_{\a} +
 \frac{h}{J_{\g}}( \hat{S}_{0} \hat{S}_{\g} +  \hat{S}_{\g} \hat{S}_{0}), \quad
  \hat{S}_{\g}  \hat{S}_{0}  -   \hat{S}_{0}\hat{S}_{\g}+ \frac{(J_{\a}-J_{\b})}{h} (
\hat{S}_{\a} \hat{S}_{\b} +  \hat{S}_{\b} \hat{S}_{\a})+    ( {l}_{\a}
\hat{S}_{\b} -  {l}_{\b} \hat{S}_{\a}) \rangle .
\end{equation}
The ideal $\cJ$ is a partial case of the ideal $\cI$
(\ref{QComReQuadZhuk}), corresponding to the following choice of the parameters
\begin{equation}\label{zcon}
  A_{\a}=\frac{h}{J_{\a}},\  B_{\a}=\frac{{(J_{\b}-J_{\g})}}{h}, \ K_{\a}=0, \  M_{\a}=0,\   N_{\a}=0, \qquad \a\in\overline{1,3}.
\end{equation}
 The parametrisation (\ref{zcon}) satisfy the condition (\ref{QCondZhuk1}), (\ref{QCondZhuk2}) of our Theorem \ref{PBWconds}, and therefore the quantum algebra   of Levin, Olshanetsky and Zotov possess   PBW property up to monomials of the third order.

The  Heisenberg equations with the Hamiltonian $\hat{S}_0$ in \cite{Zotov} are not equivalent to the dynamical system
(\ref{QCompZhuk}) on the free algebra $\mathcal{A}=\bbC\langle \hat{S}_0,\ldots,\hat{S}_3\rangle$, in other words, the coefficients $B_\a$ do not have the form (\ref{B2}), since  the ideal (\ref{LOZideal}), (\ref{zcon}) is not invariant with respect to the non-Abelian dynamics (\ref{QCompZhuk}). The quantisation presented in \cite{Zotov} can be regarded as a simultaneous deformation of both the commutative algebra of functions on the phase space and  the constants $j_{\a}$ of the dynamical system (\ref{CompZhuk0}). This deformation depends on a single parameter $\hbar$
\begin{equation}\label{jzotov}
J_{\a}= \frac{\sqrt{ {(1+h^2 j_{\b}})(1+h^2j_{\g})}}{\sqrt{(1+h^2 j_{\a})}}, \qquad \a\in \overline{1,3}.
\end{equation}
where $h^2=1/\wp(i\hbar)$  with $g_2=- 4(j_1j_2+j_2j_3+j_3
j_1),\ g_3=- 4 j_1j_2j_3$, and $j_1+j_2+j_3=0$.
In the classical limit $\hbar\to 0$ we get
\begin{equation*}
J_{\a}=1 + \hbar^2 j_{\a}+ \cO(\hbar^4), \qquad \a\in \overline{1,3},
\end{equation*}
the commutation relations yield   the   quadratic Poisson structure (\ref{qpbZhukGen}) with $c=1$, $a=b=d=0$, and the limiting system coincides with (\ref{CompZhuk0}).

 \section{Conclusion and Discussion}

The results of this paper pose several interesting mathematical and physical problems. The quantisation obtained depends on five parameters, one of which can be identified with the Planck constant. This quantisation is a generalisation of the commonly used deformation of the $so(3)$ standard Poisson bracket and Sklyanin's quadratic Poisson structure in the case of the Euler top. In the classical limit, it results in a four-parametric family of Poisson brackets, which lead to the same dynamical system as the Zhukovsky-Volterra (and Euler) top, thereby yielding identical dynamics.

In the quantum case, the problem is more subtle. Although the equations of motion formally coincide, the observables $\hat{S}_\a$ satisfy commutation relations that essentially depend on a choice of the quantisation parameters. We have reasons to believe that the resulting spectrum of the Hamiltonian also depends on the choice of the parameters. In order to compare our results with experimental data, it is necessary to develop a representation theory for the obtained five-parametric algebra and solve the spectral problem for the corresponding Hamiltonian.

\paragraph*{Acknowledgements} The authors are grateful to E. Sklyanin, N. Reshetikhin, D. Gurevich, and A. Zotov for their valuable discussions. The work of the second author was made possible by an Isaac Newton Institute and London Mathematical Society Solidarity grant, for which he expresses his gratitude. This article is partially based on work from COST Action CaLISTA CA21109, supported by COST (European Cooperation in Science and Technology), www.cost.eu.

\end{document}